\documentstyle[epsfig,12pt]{elsart} 

\begin{document}
\newcommand{\evl}{e^{\frac{i\stackrel{\leftarrow}{\partial}_{\beta}v^{\beta}+m}{\Lambda}}}
\newcommand{\evr}{e^{\frac{-v^{\beta}i\stackrel{\rightarrow}{\partial}_{\beta}+m}{\Lambda}}}
\newcommand{\esl}{e^{\frac{i\stackrel{\leftarrow}{\partial}_{\beta}v^{\beta}+m}{\Lambda}}}
\newcommand{\esr}{e^{\frac{-v^{\beta}i\stackrel{\rightarrow}{\partial}_{\beta}+m}{\Lambda}}}

\newcommand{\Gv}{\frac{g_{\omega}}{\Lambda}}
\newcommand{\Gs}{\frac{g_{\sigma}}{\Lambda}}

\newcommand{\evnml}{e^{\frac{E}{\Lambda}}}
\newcommand{\esnml}{e^{\frac{E}{\Lambda}}}

\newcommand{\evnm}{e^{-\frac{E-m}{\Lambda}}}
\newcommand{\esnm}{e^{-\frac{E+m}{\Lambda}}}

\newcommand{\Gva}{\frac{g_{\omega}}{\Lambda}}
\newcommand{\Gsa}{\frac{g_{\sigma}}{\Lambda}}

\newcommand{\Gvb}{\frac{g_{\omega}}{ (\Lambda)^{2}}}
\newcommand{\Gsb}{\frac{g_{\sigma}}{ (\Lambda)^{2}}}

\begin{frontmatter}

\title{Non-Linear Derivative Interactions in Relativistic Hadrodynamics}

\author{T. Gaitanos}, 
\author{M. Kaskulov}, 
\author{U. Mosel}
\address{
Institut f\"ur Theoretische Physik,
Universit\"at Giessen, Germany
}

\begin{abstract}
The Lagrangian density of Relativistic Hadrodynamics (RHD) 
is extended by introducing non-linear derivative (NLD) interactions of the nucleon 
with the meson fields. As a consequence, the nucleon selfenergy 
becomes both momentum and density dependent. 
With a single cut-off parameter, which regulates the NLD Lagrangian, 
our approach is compatible with results 
from microscopic nuclear matter calculations as well as with Dirac 
phenomenology. It also reproduces the correct behavior of the real part of 
the proton-nucleus potential both at low and at high proton energies. 
\end{abstract}

\begin{keyword}
relativistic hadrodynamics, non-linear derivative model, nuclear matter,
equation of state, Schr\"{o}dinger equivalent optical potential\\
PACS numbers: 21.65.-f, 21.65.Mn, 25.40.Cm 
\end{keyword}
\end{frontmatter}

\date{\today}

\section{\label{sec1}Introduction}

Relativistic mean-field models have been widely accepted as a successful 
tool for the theoretical description of different nuclear systems such as 
nuclear matter, finite nuclei and heavy-ion collisions \cite{QHD97}. The major 
advantage of relativistic mean-field (RMF) models has been a novel description 
of the saturation mechanism and an explanation of the strong spin-orbit force. 
An energy dependence of the Schr\"{o}dinger equivalent optical potential 
\cite{hama} is naturally included as a consequence of a relativistic description. 

The common starting basis for RMF models is the relativistic hadrodynamics (RHD) with 
interacting hadrons, i.e., nucleons and mesons, as the relevant degrees of freedom. 
Starting from the pioneering work of Duerr \cite{duerr}, 
simple Lagrangians have been introduced \cite{wal74}, and since 
then many different extensions and treatments have evolved. The Lagrangian 
of RHD can be treated in different approximations. 
For a practical application to nuclear matter and finite nuclei one usually 
employs the mean-field approximation. In this 
approach the nucleon selfenergies become simple functions 
of various nucleon densities, but do not explicitly depend on energy or momentum 
of the nucleon. As a consequence a linear energy dependence of the optical potential 
arises. In principle, one can go beyond the mean-field approximation in a quantum 
field theoretical treatment by a systematic diagrammatic expansion. In this case, 
the nucleon selfenergies show a much more complicated structure that 
takes the non-trivial effects of the nuclear medium into account. In 
Dirac-Brueckner (DB) calculations of nuclear matter based on a given nucleon-nucleon 
interaction it is possible to extract the corresponding nucleon selfenergies \cite{DB}. 
They depend on both the density of the nuclear medium and the energy 
and momentum of the nucleon. They reproduce the empirical saturation point of 
nuclear matter as well as the energy dependence of the optical potential for energies 
up to $\sim 350$ MeV. 

However, for practical applications the DB approach is not suitable 
due to its high complexity and its apparent limitations at high energies and densities. 
Thus, as an alternative approach to these 
{\it ab-initio} strategies for the nuclear many-body system a phenomenological 
treatment of the problem in the spirit of the mean-field approximation has been 
established. 
The basic Lagrangian of RHD has to be modified for a quantitative description of static nuclear 
systems such as nuclear matter and/or finite nuclei. Therefore, it becomes 
mandatory to introduce new terms, e.g., including non-linear selfinteractions of the 
scalar \cite{boguta} and vector \cite{sugahara} meson fields, 
or to modify existing contributions 
in the Lagrangian, e.g., introducing density dependent meson-nucleon 
couplings \cite{Toki_DDH,DDH}. The model parameters have to be fitted to properties 
of nuclear matter and/or atomic nuclei, since they cannot be derived in a simple manner 
from a microscopic description. 

For a quantitative description of dynamical systems, e.g., hadron- and heavy-ion-induced 
reactions, the energy and momentum dependence of the nuclear mean-field becomes important. 
In conventional RMF models the meson-nucleon couplings are independent of the nucleon energy 
and momentum. This leads to a linear energy dependence of the Schr\"{o}dinger equivalent 
optical potential with an unrealistically divergent behaviour at high energies. 
Indeed, analyses of proton-nucleus scattering data \cite{hama} show that the 
proton-nucleus optical potential levels off at proton energies of about $700-800$ MeV. 
Thus, alternative RMF approaches have been 
developed by including additional non-local contributions, 
i.e., by introducing Fock-terms, on the level of the RMF selfenergies 
leading to a density and energy dependence \cite{cass}. However, such a 
treatment is not covariant and also its numerical 
realization in actual transport calculations is difficult \cite{cass}. 
Another approach has been proposed in \cite{ZM} and more recently in \cite{DC} by 
introducing linear order derivative couplings in the Lagrangian of RHD. 
These additional terms lead 
to an energy (or momentum) dependence of a nucleon inside the nuclear matter, 
and modify also the 
nuclear matter equation of state (EoS). Zimanyi and Moszkowski \cite{ZM} studied such 
gradient terms with the conclusion of a softening of the nuclear EoS. A more detailed study 
by Typel \cite{DC} involved the investigation of both, the density dependence of the 
nuclear EoS and the energy dependence of the optical potential. While {\it linear}
derivative interactions of meson fields with nucleons explain the empirical energy dependence of 
the optical potential, a soft EoS results from an introduction of an explicit density dependence 
of the nucleon-meson couplings with additional parameters.

The purpose of the present work is to develop a covariant model, which 
generates both, the correct density and momentum dependence of the selfenergies 
in an unified framework. Furthermore, we find that the 
density and the momentum dependencies are indispensable and strongly correlated. 
The model proposed here is simple in realization and respects all the underlying 
symmetries of the RHD Lagrangian. The approach followed here is to extend the original 
Lagrangian of RHD \cite{wal74} by including {\it non-linear} derivative 
interactions of meson fields with nucleons.
We refer it to as non-linear derivative (NLD) model, and it is presented 
in the Section 2. The NLD Lagrangian (Section \ref{sec2}) contains not only the 
nucleon spinor and its first derivative, but it depends on all higher order derivatives 
of the nucleon field. As a consequence, the standard canonical formalism has to be generalized 
(Section \ref{subsec3}). Field equations for nucleons and meson fields can be then derived 
(Section \ref{subsec4}). The NLD model gives field equations with very simple structure 
in nuclear matter (Section \ref{sec3}) with density and energy dependent nucleon 
selfenergies. As a major advantage, both the equation of state (EoS) (density 
dependence) and the optical potential (energy dependence) are quantitatively well 
reproduced with a single parameter and the results are comparable with microscopic 
DB models, as in detailed discussed in Section \ref{sec4}. Finally, in Section \ref{sec5}, 
conclusions and an outlook complete the paper.

\section{\label{sec2}General formalism}
This section describes the details of the formalism. In the first 
subsection the general Lagrangian density is constructed and discussed. 
In the second chapter various limiting cases are investigated, before 
the Euler-Lagrange equations for the relevant degrees of freedom are 
formulated.

\subsection{\label{subsec1}The NLD Lagrangian}

The starting point is the Lagrangian density of RHD 
(RHD)~\cite{wal74}, which describes the interaction of nucleons through exchange 
of auxiliary meson fields (Lorentz-scalar, $\sigma$, and the Lorentz-vector 
isoscalar and isovector meson fields $\omega^{\mu}$ and $\vec{\rho\,}^{\mu}$) 
in the spirit of the One-Boson-Exchange (OBE) model, see for instance~\cite{DB}. 
In conventional RHD the interaction fields couple to the spinors via 
the corresponding Lorentz-density operators, e.g., 
$\overline{\Psi}\Psi\sigma$  and 
$\overline{\Psi}\gamma^{\mu}\Psi\omega_{\mu}$
($\overline{\Psi}\gamma^{\mu}\vec{\tau}\Psi\vec{\rho}_{\mu}$) 
for the scalar and vector sectors, respectively. 
In the mean-field approach to RHD the resulting mean-field potentials are
only density dependent. For instance, its Lorentz-vector part rises linearly 
with nucleon density.

In RHD the relevant degrees of freedom are nucleons and mesons ($\sigma$, $\omega$, $\rho$) 
characterized by the corresponding free Lagrangians and an interaction part between 
them:
\begin{eqnarray}
{\cal L}& = & \frac{1}{2}
\left[
	\overline{\Psi}i\gamma_{\mu}\partial^{\mu}\Psi
	- 
	(i\partial^{\mu}\overline{\Psi}) \gamma_{\mu}\Psi
\right]
- \overline{\Psi}\Psi m
\nonumber\\
& - & 
\frac{1}{2}m^{2}_{\sigma}\sigma^{2}
+\frac{1}{2}\partial_{\mu}\sigma\partial^{\mu}\sigma
+\frac{1}{2}m^{2}_{\omega}\omega_{\mu} \omega^{\mu} 
-\frac{1}{4}F_{\mu\nu}F^{\mu\nu}
\nonumber\\
& + &\frac{1}{2}m^{2}_{\rho}\vec{\rho\,}_{\mu}\vec{\rho\,}^{\mu} 
-\frac{1}{4}\vec{G\,}_{\mu\nu}\vec{G\,}^{\mu\nu}
+ {\cal L}_{int}
\;. \label{NDC-free}
\end{eqnarray}
The first line in Eq. (\ref{NDC-free}) gives the (symmetrized) free Lagrangian for the 
nucleon field $\Psi=(\Psi_{p},\Psi_{n})^{T}$ with bare mass $m$. The other terms contain the standard Lagrangians 
for the scalar and vector mesons, $\sigma$, $\omega^{\mu}$ and $\vec{\rho\,}^{\mu}$, with the 
strength tensors $F^{\mu\nu}=\partial^{\mu}\omega^{\nu}-\partial^{\nu}\omega^{\mu}$ 
and 
$\vec{G\,}^{\mu\nu}=\partial^{\mu}\vec{\rho\,}^{\nu}-\partial^{\nu}\vec{\rho\,}^{\mu}$ 
for the isoscalar and isovector meson fields, respectively. 

We propose an extension of the standard RHD by introducing non-linear derivative 
couplings into the interaction Lagrangian density. 
Non-linear derivatives on the level of the Lagrangian are introduced by imposing 
an explicit dependence of the interaction Lagrangian density on higher order partial 
derivatives of the nucleon field $\Psi$:
\begin{equation}
\label{Lf}
{\cal L}_{int} \equiv 
{\cal L}(\Psi, \, \partial_{\alpha}\Psi, 
\,\, \partial_{\alpha}\partial_{\beta}\Psi, 
\,\, \cdots, 
\,\, \overline{\Psi}, 
\,\, \partial_{\alpha}\overline{\Psi}, 
\,\, \partial_{\alpha}\partial_{\beta}\overline{\Psi}, 
\,\, \cdots)
\;.
\end{equation}
with the specified (symmetrized) interaction given by:
\begin{eqnarray}
{\cal L}_{int} & = &
\frac{g_{\sigma}}{2}
	\left[
	\overline{\Psi}
	\, \stackrel{\leftarrow}{{\cal D}}
	\Psi\sigma
	+\sigma\overline{\Psi}
	\, \stackrel{\rightarrow}{{\cal D}}
	\Psi
	\right]
\nonumber\\
& - & \frac{g_{\omega}}{2}
	\left[
	\overline{\Psi}
	 \, \stackrel{\leftarrow}{{\cal D}}
	\gamma^{\mu}\Psi\omega_{\mu}
	+\omega_{\mu}\overline{\Psi}\gamma^{\mu}
	\, \stackrel{\rightarrow}{{\cal D}}
	\Psi
	\right]
\nonumber\\
& - & \frac{g_{\rho}}{2}
	\left[
	\overline{\Psi}
	 \, \stackrel{\leftarrow}{{\cal D}}
	\gamma^{\mu}\vec{\tau}\Psi\vec{\rho}_{\mu}
	+\vec{\rho}_{\mu}\overline{\Psi}\vec{\tau}\gamma^{\mu}
	\, \stackrel{\rightarrow}{{\cal D}}
	\Psi
	\right]
\;. 
\label{NDC}
\end{eqnarray}
The interaction between the spinor fields 
$\overline{\Psi}$, $\Psi$ and the meson fields has a similar 
functional form as in standard RHD \cite{wal74}. However, now new operators ${\cal D}$ 
acting on the nucleon fields appear, which is a generic non-linear function 
of partial derivatives. The particular form of the non-linear derivative operator 
is not unique. Here a simple exponential form is assumed:
\begin{equation}
\stackrel{\rightarrow}{{\cal D}}
:= \exp{\left(\frac{-v^{\beta}i\stackrel{\rightarrow}{\partial}_{\beta}+m}{\Lambda}\right)}
\quad\mbox{, }\quad
\stackrel{\leftarrow}{{\cal D}}
:= \exp{\left(\frac{i\stackrel{\leftarrow}{\partial}_{\beta}v^{\beta}+m}{\Lambda}\right)}
\,. \label{ope}
\end{equation}
The non-linear operators (\ref{ope}) explicitely depend on the partial derivate which 
is contracted with a dimensionless unit $4$-vector $v^{\mu}$. The choice of the vector 
field $v^{\mu}$ is not fixed from first principles. It can be identified with 
the nucleon $4$-velocity $u^{\mu}$:
\begin{equation}
v^{\mu} \equiv u^{\mu} := \frac{j^{\mu}}{\sqrt{j_{\alpha}j^{\alpha}}}
\label{unitvector}
\quad ,
\end{equation}
or with some other vector meson field:
\begin{equation}
v^{\mu} := \frac{\omega^{\mu}}{\sqrt{\omega_{\alpha}\omega^{\alpha}}}
\,. \label{unitvector2}
\end{equation}
No further specification is in this context necessary. We will indeed see that 
rearrangement contributions generated by $v^{\mu}$ in the corresponding field equations 
cancel in nuclear matter. We thus consider it as an auxiliary unit vector. 
Furthermore, the additional factor $e^{m/\Lambda}$ has been introduced for convenience 
so that one does not have to renormalize the conventional meson-nucleon couplings 
$g_{\sigma,\omega,\rho}$ 
of the RHD Lagrangian. The non-linear derivative operator ${\cal D}$ contains $\Lambda$ which 
will be interpreted as a cut-off parameter. Its value is supposed to be of
natural hadronic scale of around 1 GeV. 

The entire interaction part of the Lagrangian density is symmetrized in order to 
keep the whole Lagrangian hermitian. It is invariant under global phase 
transformations which guarantees baryon number conservation. Translational invariance 
leads to a conserved energy-momentum tensor.

\subsection{\label{subsec3}Generalized formalism}

Before going forward with the field equations it is useful to study particular 
limiting cases of the general Lagrangian (\ref{NDC}) for the nucleons. The 
isovector sector is not considered here for simplicity. A series expansion 
is here useful for studying limiting cases ($i=\sigma,~\omega,~\rho$):
\newpage
\begin{eqnarray}
g_{i}e^{\frac{-v^{\beta}i\stackrel{\rightarrow}{\partial}_{\beta}}{\Lambda}} 
& = & g_{i} 
\sum_{n=0}^{\infty}~
	\frac{(-v^{\alpha}i\stackrel{\rightarrow}{\partial}_{\alpha}/\Lambda)^{n}}{n!} 
\nonumber\\
& = & 
g_{i} 
	- \frac{g_{i}}{\Lambda}v^{\alpha}
	  i\stackrel{\rightarrow}{\partial}_{\alpha}
	+ \frac{1}{2!}\frac{g_{i}}{\Lambda^{2}}
		v^{\alpha}v^{\beta}
		i\stackrel{\rightarrow}{\partial}_{\alpha}i\stackrel{\rightarrow}{\partial}_{\beta}
                \pm \cdots \nonumber 
\label{exp}
\;.
\end{eqnarray}
The interaction terms of 
the Lagrangian density (\ref{NDC}) in powers of the partial derivatives 
$i\partial^{\mu}$ takes the form
\begin{equation}
{\cal L}_{int} = {\cal L}^{(0)}_{int}+{\cal L}^{(1)}_{int}
+{\cal L}^{(2)}_{int} +\cdots
\label{L-exp}
\;.
\end{equation}

In {\it zeroth order} the interaction Lagrangian is
\begin{equation}
{\cal L}^{(0)}_{int} =  
- \frac{1}{2}g_{\omega}
	\left( 
	\overline{\Psi}
	\gamma^{\mu}\Psi\omega_{\mu}
	+
	\omega_{\mu}\overline{\Psi}
	\gamma^{\mu}\Psi
	\right)
	+
\frac{1}{2}g_{\sigma}
	\left( 
	\overline{\Psi}\Psi\sigma
	+
	\sigma\overline{\Psi}\Psi
	\right)
\;. \label{NDC2-0}
\end{equation}
Up to zeroth order in the expansion of the non-linear derivative 
terms the standard Lagrangian of RHD is therefore retained:
\begin{equation}
{\cal L}_{RHD} = \frac{1}{2}
\left[
	\overline{\Psi}\gamma_{\mu}iD^{\mu}\Psi
	+ 
	(\overline{\Psi iD^{\mu}}) \gamma_{\mu}\Psi
\right]
- \overline{\Psi}\Psi (m-g_{\sigma}\sigma)
\; ,
\end{equation}
with minimal meson-nucleon coupling terms and introducing the covariant 
derivatives $iD^{\mu}:=i\partial^{\mu}-g_{\omega}\omega^{\mu}$. 
Our general formalism contains as a limiting case, i.e., 
$\Lambda\rightarrow\infty$, the conventional RHD. 

The consideration of higher orders leads to modified Lagrangians. E.g., up to 
first order a linear dependence on these derivatives is obtained, and so 
forth. However, a truncation of the expansion to an arbitrary order is not obvious. 
Instead, the most general case of a {\it resummation of all higher 
non-linear orders} in the interaction Lagrangian is meaningful, as introduced 
in Eq. (\ref{NDC}). For this reason the full formalism will be considered for the 
rest of this work.

The NLD Lagrangian $\mathcal{L}$ is a functional of not only $\Psi$, $\overline{\Psi}$ and 
their first derivatives, but it depends on all higher order space-time derivatives 
of the spinor fields $\Psi$ and $\overline{\Psi}$.
Thus the standard expressions for the Euler-Lagrange equations and the Noether Theorem 
do not apply. Instead of, we have to explicitly derive the generalized
Noether Theorem for a 
Lagrangian density which contains higher order derivatives, e.g., 
${\cal L} = {\cal L}(\phi, \partial_{\alpha}\phi, \partial_{\alpha}\partial_{\beta}\phi, \cdots)$. 
This step is important in defining conserved quantities such as the energy-momentum tensor 
and Noether currents. 

For a general functional containing fields and their higher order derivatives the Euler-Lagrange 
equations take the form:
\begin{eqnarray}
\frac{\partial{\cal L}}{\partial\phi}
-
 \partial_{\alpha}\frac{\partial{\cal L}}{\partial(\partial_{\alpha}\phi)}
+
 \partial_{\alpha}\partial_{\beta}\frac{\partial{\cal L}}{\partial(\partial_{\alpha}\partial_{\beta}\phi)}
&& + \cdots  + \\
&& (-)^{n}\partial_{\alpha_{1}}\partial_{\alpha_{2}}\cdots\partial_{\alpha_{n}}
\frac{\partial{\cal L}}
{\partial(\partial^{\alpha_{1}}\partial^{\alpha_{2}}\cdots\partial^{\alpha_{n}})}=
0 \nonumber
\;. \label{Euler}
\end{eqnarray}

For a general transformation of the fields and the coordinates the variational 
method for an arbitrary Lagrangian density (\ref{Lf}), which contains 
higher order derivatives, leads to the following continuity equation (see 
appendix \ref{app1} for detailed derivations):
\begin{equation}
\partial_{\mu}
\left[
  J^{\mu} - T^{\mu\nu}\delta x_{\nu}
\right] = 0
\;. \label{Noether-a}
\end{equation}
where $J^{\mu}$ and $T^{\mu\nu}$ are defined as:
\begin{eqnarray}
J^{\mu} & = &
  \left[
    \frac{\partial{\cal L}}{\partial(\partial_{\mu}\phi)}
  - \partial_{\beta}
    \frac{\partial{\cal L}}{\partial(\partial_{\mu}\partial_{\beta}\phi)}
  + \partial_{\beta}\partial_{\gamma}
    \frac{\partial{\cal L}}{\partial(\partial_{\mu}\partial_{\beta}\partial_{\gamma}\phi)}
  \mp \cdots
  \right]\delta\phi_{T}
\nonumber\\
& + & \left[
    \frac{\partial{\cal L}}{\partial(\partial_{\mu}\partial_{\beta}\phi)}
  - \partial_{\gamma}
    \frac{\partial{\cal L}}{\partial(\partial_{\mu}\partial_{\beta}\partial_{\gamma}\phi)}
  \pm \cdots
  \right]\partial_{\beta}\delta\phi_{T}
\nonumber\\
& + & \left[
    \frac{\partial{\cal L}}{\partial(\partial_{\mu}\partial_{\nu}\partial_{\xi}\phi)}
  - \partial_{\gamma}
    \frac{\partial{\cal L}}{\partial(\partial_{\mu}\partial_{\gamma}\partial_{\nu}\partial_{\xi}\phi)}
  \pm \cdots
  \right]\partial_{\nu}\partial_{\xi}\delta\phi_{T}
\nonumber \\
& + & \cdots \nonumber\\ 
\label{Noether-Current}\\
T^{\mu\nu} & = &
  \left[
    \frac{\partial{\cal L}}{\partial(\partial_{\mu}\phi)}
  - \partial_{\beta}
    \frac{\partial{\cal L}}{\partial(\partial_{\mu}\partial_{\beta}\phi)}
  + \partial_{\beta}\partial_{\gamma}
    \frac{\partial{\cal L}}{\partial(\partial_{\mu}\partial_{\beta}\partial_{\gamma}\phi)}
  \mp \cdots
  \right]\partial^{\nu}\phi
\nonumber\\
& + & \left[
    \frac{\partial{\cal L}}{\partial(\partial_{\mu}\partial_{\beta}\phi)}
  - \partial_{\gamma}
    \frac{\partial{\cal L}}{\partial(\partial_{\mu}\partial_{\beta}\partial_{\gamma}\phi)}
  \pm \cdots
  \right]\partial_{\beta}\partial^{\nu}\phi
\nonumber\\
& + & \left[
    \frac{\partial{\cal L}}{\partial(\partial_{\mu}\partial_{\beta}\partial_{\xi}\phi)}
  - \partial_{\gamma}
    \frac{\partial{\cal L}}{\partial(\partial_{\mu}\partial_{\gamma}\partial_{\beta}\partial_{\xi}\phi)}
  \pm \cdots
  \right]\partial_{\beta}\partial_{\xi}\partial^{\nu}\phi 
+  \cdots
-  g^{\mu\nu}{\cal L}
\;. \label{Noether}
\end{eqnarray}

In case of Poincare' transformations, i.e., 
$\delta\phi_{T}=\phi^{\prime}(x^{\prime})-\phi(x)=0$ and $x^{\prime}=x^{\mu}+\delta x^{\mu}$ 
we obtain $\partial_{\mu}T^{\mu\nu}=0$ and therefore the conserved 
energy-momentum tensor density. In case of global phase transformations, i.e., 
$\delta\phi_{T}\approx i\epsilon\phi$ and $\delta x^{\mu}=0$, the continuity equation 
$\partial_{\mu}J^{\mu}=0$ is retained.

Contrary to the standard 
Noether theorem, now more terms proportional to higher order derivatives of the 
variation of the field $\delta\phi$ appear, and for each order with respect to the 
derivative on $\delta\phi$, an infinite series appears. However, we will see that 
the infinite series can be resummed to compact expressions and that 
all the equations will simplify considerably.

\subsection{\label{subsec4}Field equations for all degrees of freedom}

The application of the generalized Euler-Lagrange equations (\ref{Euler}) to the full 
Lagrangian density (\ref{NDC}) with respect to the Spinor field $\Psi$ leads to a 
Dirac equation with selfenergies, which in general contain infinite series with respect to 
the partial derivatives 
\footnote{In order to keep the presentation of the model and the discussions 
transparent, we omit the $\rho$-meson contributions in all the 
expressions. It is, however, included in the calculations for neutron 
matter.
}. As shown in Appendix \ref{app2}, all series can be 
resummed to simple exponential functions leading to the following Dirac equation:
\begin{equation}
\left[
	\gamma_{\mu}(i\partial^{\mu}-\Sigma^{\mu}) - 
	(m-\Sigma_{s})
\right]\Psi = 0
\;, 
\label{Dirac}
\end{equation}
with Lorentz-vector and Lorentz-scalar selfenergies defined as:
\begin{equation}
\Sigma^{\mu} = g_{\omega}\omega^{\mu}\evr + \Sigma^{\mu}_{r}
\quad\mbox{, } \\ \quad
\Sigma_{s} = g_{\sigma}\sigma\esr
\;. \label{Sigma}
\end{equation}
Both vector and scalar selfenergies $\Sigma^{\mu},~\Sigma_{s}$ show an explicit linear 
behavior with respect to the corresponding fields, $\omega^{\mu}$ and $\sigma$, 
respectively, which looks like as in the linear Walecka model of RHD. It has been widely discussed 
in the literature that the linear Walecka model does not reproduce the compressibility of 
nuclear matter at saturation density, except if selfinteractions between 
the $\sigma$ meson fields \cite{wal74,boguta} and the $\omega$ fields \cite{sugahara} are 
introduced. One may thus expects the appearance of the 
same problem also in our formalism. On the other hand, the meson fields will contain a 
residual non-linear density behavior, which will be discussed below, together with the 
rearrangement term $\Sigma^{\mu}_{r}$. Apart from the density dependence, 
the selfenergies contain now a non-linear dependence on the partial derivatives, 
which in nuclear matter are related to the $4$-momentum of a moving nucleon inside 
the nuclear matter at rest.

Proca and Klein-Gordon equations for the meson fields are immediately derived 
from the Lagrangian density (\ref{NDC}) by considering the explicit 
$\omega$-dependence of the $4$-vector $v^{\mu}$ (see again Eq. (\ref{unitvector2})):
\begin{eqnarray}
\partial_{\mu}F^{\mu\nu} + m_{\omega}^{2}\omega^{\nu} &=& 
\frac{1}{2}g_{\omega}
\left[
	\overline{\Psi}\evl \gamma^{\nu}\Psi + \overline{\Psi}\gamma^{\nu}\evr \Psi
\right] + \Omega^{\nu}_{r},
\label{omega_meson} \\
\nonumber\\
\partial_{\alpha}\partial^{\alpha}\sigma + m_{\sigma}^{2}\sigma &=& 
\frac{1}{2}g_{\sigma}
\left[
	\overline{\Psi}\evl \Psi + \overline{\Psi}\evr \Psi
\right]
\;. \label{sigma_meson}
\end{eqnarray}
The Proca equation for the $\omega$ meson field (\ref{omega_meson}) shows 
a rather complex structure of the source term indicating a highly non-linear 
behavior of the $\omega$ field (the rearrangement term $\Omega_{r}^{\mu}$ 
is given in Eq. (\ref{rearr_o}), however, it is not relevant as discussed 
below). 
The same feature also holds for the source 
term of the $\sigma$-meson field, although in the latter case its structure 
is simpler. Note that the meson field equations for the two mesons are coupled 
through the non-linear exponential terms. In nuclear matter these 
equations will simplify considerably, however, the coupling between both 
meson equations will remain. This feature together with the non-linearities in the 
source terms will generate a residual non-linear density behavior also for the 
vector field, which in the standard Walecka model is linear in density. 

Depending on the particular choice of the the auxiliary unit vector 
in Eqs. (\ref{unitvector}) or (\ref{unitvector2}) 
rearrangement terms appear in the vector part 
of the nucleon selfenergy ($\Sigma^{\mu}_{r}$) or in the source term of the 
Proca equation ($\Omega^{\mu}_{r}$). An important feature is, 
however, that independent of the choice of $v^{\mu}$ the rearrangement terms 
exactly cancel in nuclear matter (see appendix \ref{app2a} and \ref{app2b}) simplifying the 
equations of motion to a large extent. We, therefore, do not specify these 
rearrangement contributions here any further.

\section{\label{sec3}Application to infinite nuclear matter}

Infinite nuclear matter is characterized by a homogeneous, 
isotropic and stationary system at zero temperature. Due to these symmetries 
the field equations simplify: the densities are constant in space-time, and 
only the $0$-component of the currents survives. Consequently, the 
meson fields are constant numbers and the spatial components of the 
Lorentz-vector $\omega$-field vanish. In the spirit of the mean-field 
approximation all meson fields are treated as classical fields implying 
the replacement of the source term operators in the meson field equations 
by their expectation values. Therefore 
$v^{\mu}=\omega^{\mu}/\sqrt{\omega_{\alpha}\omega^{\alpha}}=(1,~\vec{0})$ or 
$v^{\mu}=j^{\mu}/\sqrt{j_{\alpha}j^{\alpha}}=(1,~\vec{0})$, since the space-like 
components of the usual nucleon current $j^{\mu}=<\overline{\Psi}\gamma^{\mu}\Psi>$ 
vanish in average for nuclear matter at rest.

In nuclear matter solutions of the spinor field are found with the plane 
wave ansatz (for simplicity isospin is neglected):
\begin{equation}
\Psi(s,\vec{p}) = u(s,\vec{p})e^{-ip^{\mu}x_{\mu}}
\:, \label{plane_wave}
\end{equation}
with $s$ and $p^{\mu}=(E,\vec{p})$ being the spin and $4$-momentum of the 
nucleon, $x^{\mu}$ its time-space coordinate, and $u$ the spinor field 
for positive energy states.

In nuclear matter the rearrangement contributions cancel and 
the Dirac equation (\ref{Dirac}) maintains its form with selfenergies given by:
\begin{equation}
\Sigma_{v} = g_{\omega}\omega_{0}e^{-\frac{E-m}{\Lambda}}
\quad\mbox{, }\quad
\Sigma_{s} = g_{\sigma}\sigma e^{-\frac{E-m}{\Lambda}}
\; . \label{SelfenNM}
\end{equation}
As one can see when the nucleons are at rest, $E=m$, the exponential factor is 
equal to unity and the equations are reduced to the ones from the Walecka model. 
The selfenergies show now a transparent behavior in energy. They contain a 
non-linear energy dependence, in particular, they exponentially decrease 
with increasing energy, as also expected from Dirac Phenomenology \cite{hama}. 
They are linear in the meson fields as in the linear Walecka model, however, 
this does not necessarily imply a linear density behavior of the mean field.

Solutions of the Dirac equation are obtained by inserting (\ref{plane_wave}) 
and (\ref{SelfenNM}) into the Dirac equation (\ref{Dirac}):
\begin{equation}
\gamma_{0}E^{*}\Psi = 
\left( 
\vec{\gamma}\cdot\vec{p}+m^{*}
\right)\Psi
\; , \label{DiracNM}
\end{equation}
with 
\begin{equation}
E^{*} := E - g_{\omega}\omega_{0}e^{-\frac{E-m}{\Lambda}}
\quad\mbox{, }\quad
m^{*} := m - g_{\sigma}\sigma e^{-\frac{E-m}{\Lambda}}
\;. \label{effective}
\end{equation}
The Dirac equation (\ref{DiracNM}) has the same structure as in 
the free space, however, with modified effective energy $E^{*}$ and 
effective (Dirac) mass $m^{*}$. The spatial component of the 
$4$-momentum $\vec{p}$ remains unchanged in nuclear matter. Solutions 
of the Dirac equation (\ref{DiracNM}) are found in the usual way:
\begin{equation}
u(s,\vec{p}\,) = N
\left(
\begin{array}{c}
\phi_{s} \\
\frac{ \vec{\sigma}\cdot\vec{p}}{E^{*}+m^{*}}\phi_{s}\\
\end{array}
\right)
\; , \label{Spinor}
\end{equation}
with spin eigenfunctions $\phi_{s}$ and with the normalization 
$N=\sqrt{\frac{E^{*}+m^{*}}{2E^{*}}}$, which guarantees the orthogonality 
relation of the spinors. The following dispersion relation holds
\begin{equation}
E^{*2} - \vec{p}^{2} = m^{*2}
\;. \label{mass-shel}
\end{equation}
Note that, $E^{*}$ and $m^{*}$ depend both on density and, particular, on 
energy implying self consistency to obtain the dispersion relation 
$E=E(p)$.

In nuclear matter the NLD equations of motion for $\sigma$ and $\omega$ 
simplify significantly to standard meson field equations:
\begin{equation}
m_{\omega}^{2}\omega^{0} = g_{\omega}\rho_{0}
\quad\mbox{, }\quad
m_{\sigma}^{2}\sigma = g_{\sigma}\rho_{s} 
\;, \label{mesonsNM}
\end{equation}
with the following source terms ($\kappa=4,2$ for nuclear matter and neutron matter):
\begin{eqnarray}
\rho_{s} & = & \langle \overline{\Psi} e^{-\frac{E-m}{\Lambda}}\Psi\rangle 
= \frac{\kappa}{(2\pi)^{3}}\; \int 
d^{3}p \frac{m^{*}}{E^{*}} e^{-\frac{E-m}{\Lambda}}\Theta (p-p_{F}),
\nonumber\\
\rho_{0} & = & \langle \overline{\Psi} \gamma^{0} e^{-\frac{E-m}{\Lambda}}\Psi\rangle
= \frac{\kappa}{(2\pi)^{3}}\; \int d^{3}p e^{-\frac{E-m}{\Lambda}}\Theta (p-p_{F})
\;. \label{dens-1}
\end{eqnarray}
Note that, the density $\rho_{0}$ is not related to the conserved nucleon density 
$\rho_{B}$. Later on, this quantity will be specified explicitely. Both vector 
and scalar densities $\rho_{0}$ and $\rho_{s}$ differ considerably from those of 
the conventional Walecka model due to the dispersion relation (\ref{mass-shel}) 
and the appearance of the exponential terms, which act as {\it damping factors}. 
These effects lead to a coupled set of equations between the meson fields and the 
dispersion relation. They generate a residual non-linear density dependence for both 
the scalar field and especially the vector $\omega$-meson. 

According to the generalized Noether-theorem (\ref{Noether}) a conserved nucleon density 
and an equation of state can be derived (see appendix \ref{app3}). The equation 
\begin{equation}
J^{0} \equiv \rho_{B} = \langle \overline{\Psi}\gamma^{0}\Psi \rangle
+ \frac{g_{\omega}}{\Lambda}
	\langle \overline{\Psi}\gamma^{0}\evnm \Psi \rangle \omega_{0}
- \frac{g_{\sigma}}{\Lambda}
	\langle \overline{\Psi}\evnm \Psi \rangle \sigma
\label{rhoBar}
\end{equation}
characterizes the conserved nucleon density $\rho_{B}$, from which also the 
relation between the Fermi momentum $p_{F}$ and $\rho_{B}$ is uniquely determined. 
The expectation values 
$\rho_{0}=\langle \overline{\Psi}\gamma^{0}\evnm \Psi \rangle$ and 
$\rho_{s}=\langle \overline{\Psi}\evnm \Psi \rangle$ have been already evaluated in 
(\ref{dens-1}). The expectation value 
$\langle \overline{\Psi}\gamma^{0}\Psi \rangle=
\frac{\kappa}{(2\pi)^{3}}(4\pi)\frac{p_{F}^{3}}{3}$ is just the usual density of the 
Walecka model \cite{wal74}. 
Unlike the Walecka model, which is recovered in the limiting case of 
$\Lambda\to\infty$, the baryon density in the NLD approach shows a more 
complex structure. This results from the 
application of the generalized Noether-theorem to the NLD Lagrangian in nuclear 
matter, in which the non-linear derivative terms are coupled with the meson fields, 
as shown in appendix \ref{app3}. 
The extra terms in Eq. (\ref{rhoBar}) reflect the dressing of the 
nucleons with the meson clouds. The latter start to contribute at densities above 
saturation, e.g., at $\rho_{B}=2\rho_{sat}$ the nucleons contribute about $92\%$ and 
the mesons $8\%$ to the baryon density.

The Equation of State (EoS) is obtained from the $00$-component of the 
energy-momentum tensor $T^{\mu\nu}$:
\begin{eqnarray}
T^{00} \equiv  \epsilon &=& 
\langle \overline{\Psi}\gamma^{0} E \Psi \rangle
+ \frac{g_{\omega}}{\Lambda}
	\langle \overline{\Psi}\gamma^{0}\evnm E \Psi \rangle \omega_{0}
- \frac{g_{\sigma}}{\Lambda}
	\langle \overline{\Psi}\evnm E \Psi \rangle \sigma
\nonumber\\
&& +  \frac{1}{2}
\left( 
	m_{\sigma}^{2}\sigma^{2} - m_{\omega}^{2}\omega_{0}^{2}
\right)
\;, \label{eos}
\end{eqnarray}
with the additional expectation values:
\begin{eqnarray}
\rho_{s}^{E} & = & \langle \overline{\Psi}E \evnm \Psi\rangle 
= \frac{\kappa}{(2\pi)^{3}}\; \int d^{3}p \frac{m^{*}}{E^{*}} E \evnm\Theta (p-p_{F}),
\nonumber\\
\rho_{0}^{E} & = & \langle \overline{\Psi} E \gamma^{0} \evnm \Psi \rangle
= \frac{\kappa}{(2\pi)^{3}}\; \int d^{3}p E \evnm\Theta (p-p_{F})
\;. \label{dens-2}
\end{eqnarray}

\section{\label{sec4}Results}

The set of coupled equations (\ref{mass-shel},\ref{mesonsNM}) has to 
be solved self consistently together with the numerical integrations 
for the different densities (\ref{dens-1},\ref{dens-2}). This procedure 
is done as function of the Fermi momentum $p_{F}$ and of the momentum $p$. 
The relation between the nucleon density $\rho_{B}$ and the Fermi 
momentum $p_{F}$ is given by Eq. (\ref{rhoBar}).

\begin{figure}[t]
\unitlength1cm
\begin{picture}(18.,9.0)
\put(2.0,0.0){\makebox{\epsfig{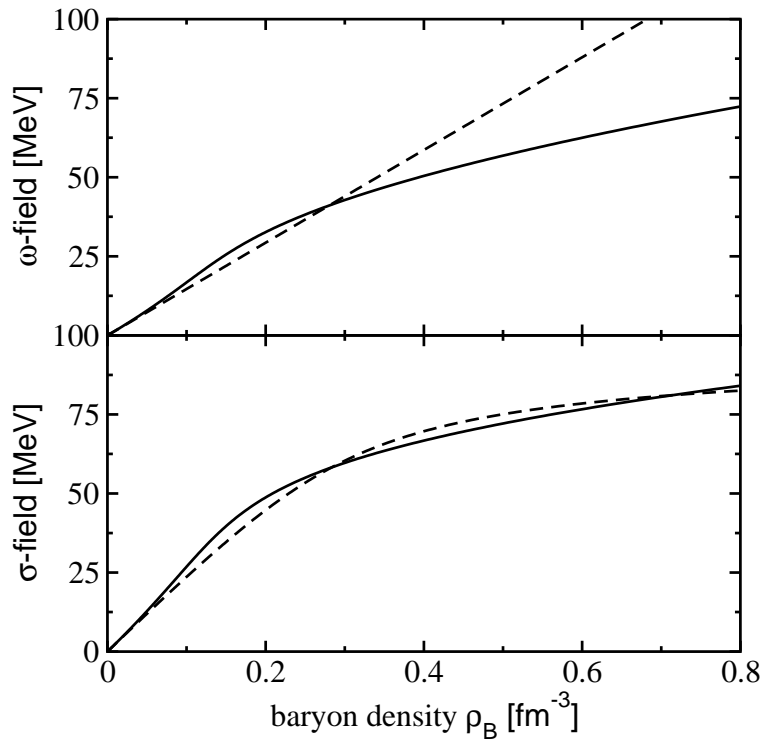}}}
\end{picture}
\caption{Vector ($\omega$, upper panel) and scalar ($\sigma$, lower panel) 
meson fields as function of the nucleon density $\rho_{B}$. 
Dashed: Linear Walecka model, solid: NLD model.
}
\label{fig1}
\end{figure}
In principle, the NLD model contains no parameters. The original 
$\sigma NN$ and $\omega NN$ couplings can be taken from any 
linear Walecka model \cite{wal74}, as it has been done here. The cut-off parameter 
$\Lambda$ has to be of natural size, i.e., of typical hadronic mass scale 
in this problem. In the following, $\Lambda=0.770$ GeV is chosen for 
a quantitative discussion of the results. Of course, different cut-off 
parameters $\Lambda_{\sigma,\omega}$ can be also used, which may be 
necessary when applying the NLD model to more complex nuclear systems 
such as finite nuclei. However, this is not the scope of the present work.

We start the discussion with the dependence of the isoscalar meson fields 
as function of the nucleon density $\rho_{B}$, see Fig. \ref{fig1}. In the 
standard linear Walecka model the $\omega$-field shows the typical linear 
density behavior, while the scalar meson field saturates with increasing 
$\rho_{B}$ due to the suppression $\gamma$-factor $m^{*}/E^{*}$. In the NLD 
model, however, a saturation of the $\omega$-field is observed too. 
This result is similar to that of Ref. \cite{sugahara}, in which 
selfinteractions of the $\omega$ meson field were considered.  
This non-linear effect on the vector meson arises from the damping factor 
$e^{-E/\Lambda}$, as discussed above. Non-linear effects to the scalar 
meson, on the other hand, are moderate due to partial compensation effects 
between the damping factor and the $m^{*}/E^{*}$ term. 

\begin{figure}[t]
\unitlength1cm
\begin{picture}(18.,9.0)
\put(2.0,0.0){\makebox{\epsfig{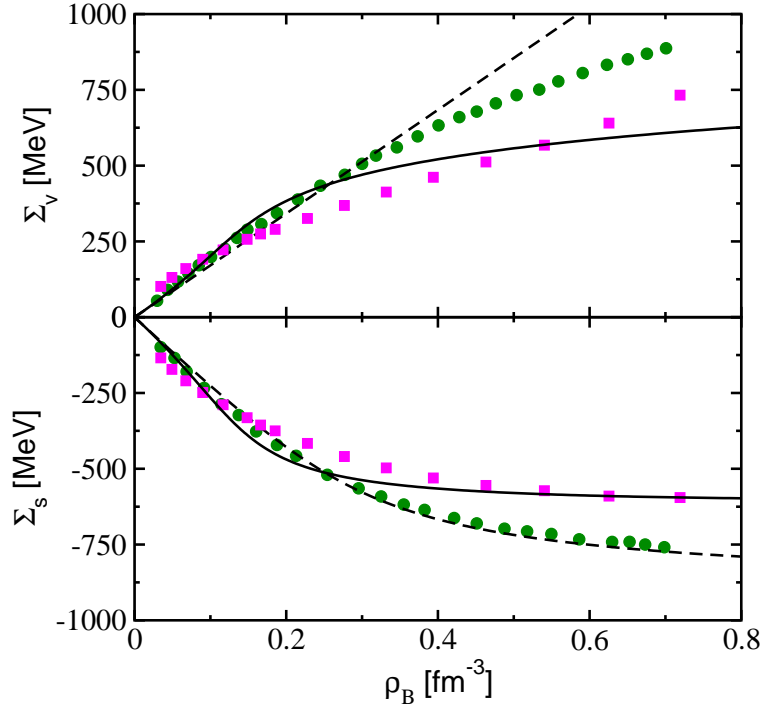}}}
\end{picture}
\caption{Vector (upper panel) and scalar (lower panel) selfenergies as function 
of nucleon density. Dashed: Linear Walecka model, solid: NLD model, 
filled squares: DB model \protect\cite{DB}, filled circles: Sugahara 
and Toki \cite{sugahara}.
}
\label{fig2}
\end{figure}
The selfenergies $\Sigma_{s,v}$ are proportional to the meson fields 
(see Eqs. (\ref{SelfenNM})), however, they are  dominated by the 
factor $e^{-E/\Lambda}$ which appears explicitly and is itself also 
density dependent through the in-medium dispersion relation. As can be seen 
in Fig. \ref{fig2}, the scalar part 
(lower panel) is thus further suppressed in the NLD model compared with the 
standard linear Walecka model. The density dependence of the vector field 
in the NLD approach differs considerably from the usual linear dependence, 
in particular, at nucleon densities above saturation 
$\Sigma_{v}$ is strongly suppressed and tends to saturate at very high densities. 
The agreement between the NLD model and  microscopic DB 
calculations \cite{DB} is remarkable over a wide region in nucleon density. 
Another possibility of suppressing the linear density dependence 
of the vector field was considered in Ref. \cite{sugahara}, in which a 
selfinteraction term of the $\omega$ meson was introduced in the 
RMF Lagrangian. As can be seen again in Fig. \ref{fig2}, both models 
are very close to each other around saturation density and below, but differ 
considerably for $\rho_{B}\geq 0.3~fm^{-3}$.

\begin{figure}[t]
\unitlength1cm
\begin{picture}(18.,10.0)
\put(0.0,0.0){\makebox{\epsfig{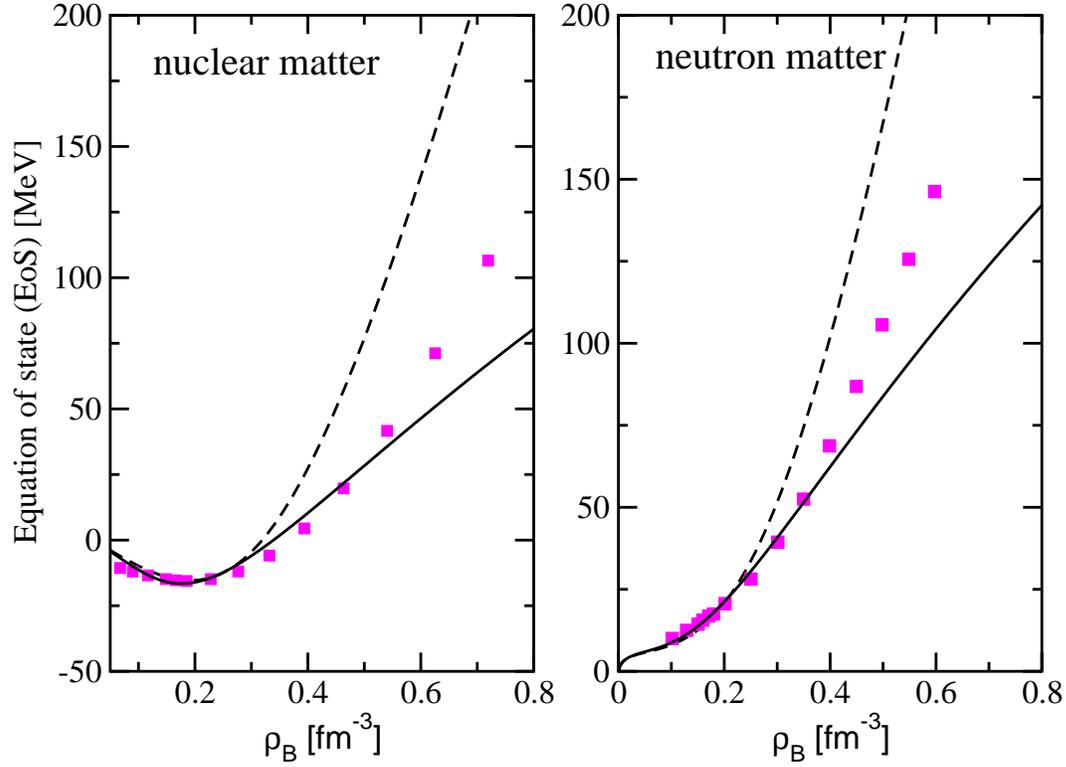}}}
\end{picture}
\caption{Equation of state for nuclear (left panel) and pure neutron matter 
(right panel). Dashed: Walecka model, solid: NLD model, 
filled squares: DB model \protect\cite{DB}.
}
\label{fig3}
\end{figure}
The non-linear density dependence significantly affects the equation of state 
(EoS), i.e., the binding energy per nucleon as function of nucleon density. 
This is demonstrated in Fig. \ref{fig3} for nuclear and pure neutron matter. 
The standard linear Walecka model leads to an EoS with high stiffness resulting 
in a very high value for the compression modulus. The NLD model weakens the 
stiffness of the EoS for nuclear and pure neutron matter to a large extent. 
The agreement of the NLD-EoS with the underlying DB theory is successful 
indicating that saturation nuclear matter properties, e.g., binding energy 
per nucleon and compression modulus at saturation density, are fairly well 
reproduced by the NLD approach. 
It thus turns out that standard linear Walecka models without any explicit 
introduction of selfinteractions of the meson fields are able to 
describe nuclear matter properties by the introduction of non-linear derivatives 
with a single cut-off parameter. 

\begin{figure}[t]
\unitlength1cm
\begin{picture}(18.,9.0)
\put(2.0,0.0){\makebox{\epsfig{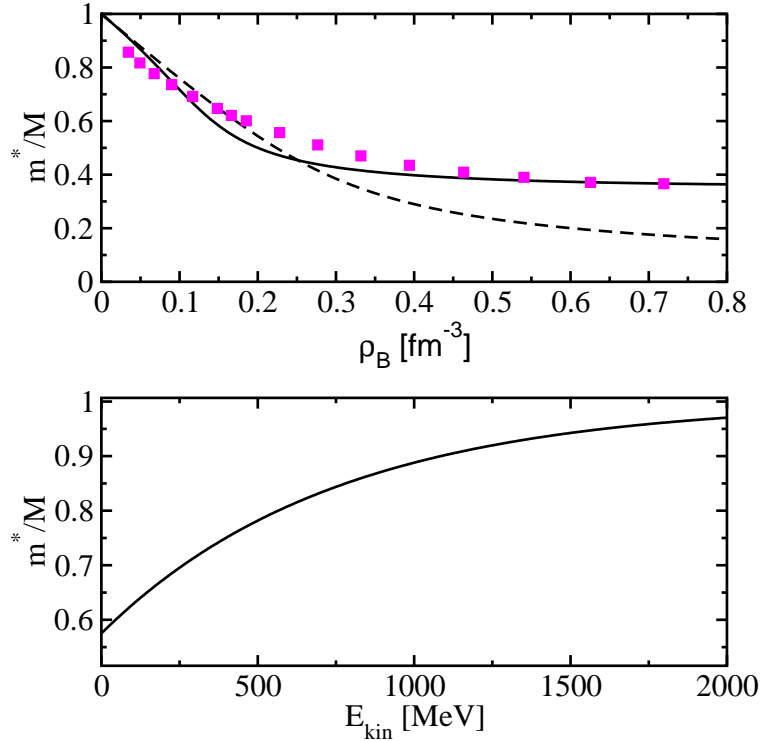}}}
\end{picture}
\caption{Upper panel: Nucleon density dependence of the effective (Dirac) mass 
$m^{*}$ in units of the bare nucleon mass $M$. Dashed: Walecka model, 
solid: NLD model, filled squares: DB model \protect\cite{DB}. 
Lower panel: Energy dependence of the same quantity in the NLD approach. 
}
\label{fig4}
\end{figure}
Similar effects are observed in the effective mass, as shown in the upper panel of 
Fig. \ref{fig4}. At saturation density a value of $m^{*}\approx 0.65 m$ is 
obtained, which is again very close to DB predictions. The increase 
of the effective (Dirac) mass at high nucleon densities ($\rho_{B}\geq 0.3~fm^{-3}$) 
in the NLD model reflects the attractive character of the mean-field, which is 
consistent with the suppression of the scalar selfenergy at high densities and 
is quantitatively compatible with microscopic DB calculations. 

An important feature of the NLD model is the prediction of an 
energy (or momentum) dependence of the nuclear mean-field with only one parameter. 
Dirac Phenomenology on elastic proton-nucleus scattering predicts a non-linear energy 
dependence of the Schr\"{o}dinger equivalent optical potential (see below), which 
cannot be reproduced in linear Walecka models nor in their extensions to non-linear 
meson field terms. The question arises if the NLD model 
can reproduce this feature with the same parameter $\Lambda$ as used for the 
density dependence. 

In the following we consider the situation of a nucleon with particular momentum 
$p$ (or kinetic energy $E_{kin}$) relative to nuclear matter at rest. 
The kinetic energy for an incident free nucleon with mass $m$ and momentum 
$p$ is usually defined as $E_{kin}=\sqrt{p^{2}+m^{2}}-m$. In the nuclear medium 
$E_{kin}$ reads \cite{DC,DB-e}
\begin{equation}
E_{kin}=E-m=\sqrt{p^{2}+m^{*2}}+\Sigma_{v}-m
\quad .
\end{equation}
As a first impression 
Fig. \ref{fig4} (panel on the bottom) shows the energy dependence of the effective (Dirac) 
mass at fixed nucleon density $\rho_{B}=\rho_{sat}=0.16~fm^{-3}$. In the standard 
linear Walecka model $m^{*}$ is independent on energy, since the scalar selfenergy 
is momentum independent. In the NLD model, however, the effective mass non-linearly 
rises with energy, and in the limit $E\rightarrow\infty$ (not shown in the figure) 
$m^{*}\rightarrow m$ is achieved, since $e^{-E/\Lambda}\rightarrow 0$. This result 
is in remarkable agreement with Dirac phenomenology \cite{hama}.

\begin{figure}[t]
\unitlength1cm
\begin{picture}(18.,9.0)
\put(2.0,0.0){\makebox{\epsfig{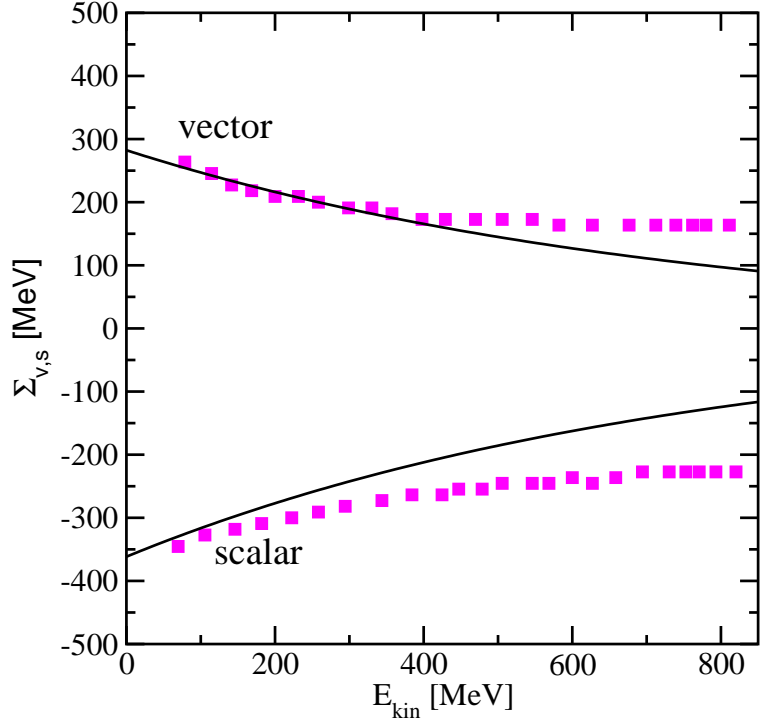}}}
\end{picture}
\caption{NLD vector and scalar selfenergies (solid curves, as indicated) 
as function of the kinetic energy at fixed saturation density. The filled 
squares show DB calculations taken from \protect\cite{DB-e}.
}
\label{fig5}
\end{figure}
The explicit energy dependence of the scalar and vector selfenergies in the 
NLD model are shown in Fig. \ref{fig5} (solid curves). 
According to Eqs. (\ref{SelfenNM}) an approximately exponential decrease of 
$\Sigma_{s,v}$ with increasing energy can be seen, which is unambiguously 
fixed by the same cut-off parameter $\Lambda$ as used for the 
density dependence. The agreement with DB results is again remarkable 
in an energy region up to $E_{kin}\simeq 0.4$ GeV. The DB model describes 
very well saturation properties of nuclear matter. It further gives a realistic energy 
dependence of the selfenergies in the considered energy region by 
using the parameters of the underlying bare $NN$-potential \cite{DB-e}. 
The deviation of the fields between the NLD 
and DB approaches for energies above $\sim 0.4$ GeV will be important in describing 
the energy dependence of the optical potential.

\begin{figure}[t]
\unitlength1cm
\begin{picture}(18.,9.0)
\put(1.0,0.0){\makebox{\epsfig{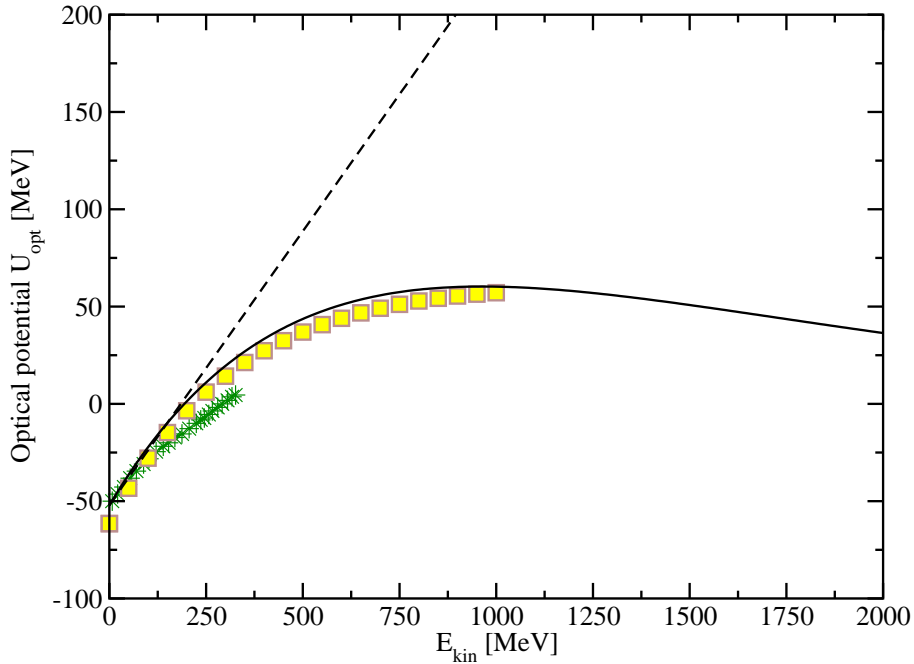}}}
\end{picture}
\caption{(left) Energy dependence of the Schr\"{o}dinger equivalent optical 
potential at density $\rho_{sat}=0.16~fm^{-3}$. The theoretical calculations 
(dashed: linear Walecka model, solid: NLD model) are compared with data from 
Dirac phenomenology (filled squares) \protect\cite{hama} and with DB 
calculations (filled stars) \protect\cite{DB-e}.
}
\label{fig6}
\end{figure}

The energy dependence of the nuclear mean-field is usually determined by 
Dirac phenomenology in elastic nucleon-nucleus scattering \cite{hama}. The 
essential physical object 
here is the Schr\"{o}dinger equivalent optical potential 
$V_{opt}$, which serves as a convenient means to characterize the in-medium 
interaction of a nucleon with momentum $p$ relative to nuclear matter at rest. 
It is obtained by rearranging the Dirac equation (\ref{DiracNM}) into a 
Schr\"{o}dinger-like equation for the large (upper) component of the nucleon 
spinor, i.e., by performing a non-relativistic reduction of the Dirac equation. 
This procedure applied to Eq. (\ref{DiracNM}) leads to a central 
Schr\"{o}dinger-like potential $V_{opt}$
\begin{equation}
 V_{\rm opt} = \frac{E}{m} \Sigma_{v} - \Sigma_{s}
 + \frac{1}{2m} \left( \Sigma^{2}_{s} - \Sigma_{v}^{2}\right)
\:. \label{V_opt}
\end{equation}
It is determined by the scalar and vector selfenergies and it rises 
linearly with energy, if the selfenergies do not depend explicitely 
on momentum. This is the case of the linear Walecka model, as 
can be seen in Fig. \ref{fig6} (dashed curve). 
The DB model (filled stars), on the other hand, reproduces the empirical behavior 
of the optical potential at low energies, since the parameters of the underlying 
free NN-interaction are fitted to low energy scattering data \cite{DB-e}.
The NLD model (solid curve) with its 
non-linear energy dependence weakens the linear stiffness of the original 
Walecka model to a large extent, and the empirical energy behaviour 
of the optical potential can be reproduced without the introduction 
of any further parameters.  

\begin{figure}[t]
\unitlength1cm
\begin{picture}(18.,9.0)
\put(1.0,0.0){\makebox{\epsfig{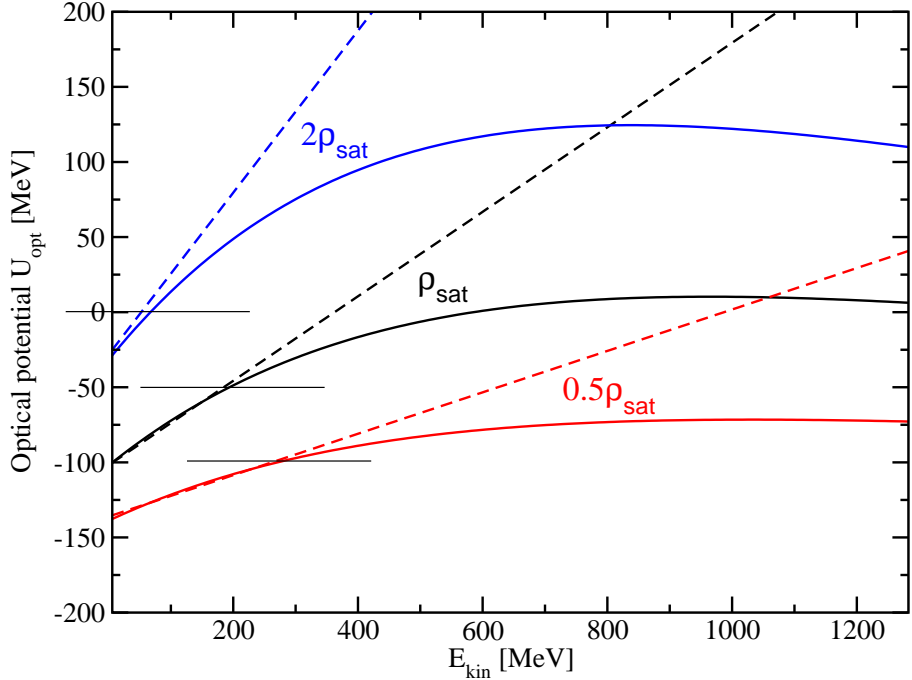}}}
\end{picture}
\caption{Energy dependence of the Schr\"{o}dinger equivalent optical 
potential at different densities, as indicated. For each density both 
calculations are shown, linear Walecka model (dashed) and NLD approach 
(solid). The potentials at $\rho_{B}=0.5\rho_{sat}$ (dotted lines), 
$\rho_{B}=\rho_{sat}$ (solid lines) and $\rho_{B}=2\rho_{sat}$ 
(dot-dashed lines) are shifted downwards by $100$ MeV, $50$ MeV and 
$0$ MeV, respectively. Horizontal lines mark the range in the energy 
where the potentials cross zero.
}
\label{fig7}
\end{figure}

The strong suppression of the mean field at higher energies at saturation 
density is maintained also at higher nucleon densities, as shown in 
Fig. \ref{fig7}. For highly compressed matter, as it occurs in high energy 
heavy ion collisions, one thus expects a significant suppression of the 
repulsion of the nuclear mean field in the NLD approach. The novel result 
of the NLD model is thus a softening of the EoS at high densities and 
the Schr\"{o}dinger equivalent optical potential at high energies and 
also high compressions. These features will be of particular interest 
for the theoretical description of high energy heavy-ion collisions, as 
they will be investigated at the compressed baryonic matter experiment at 
the new FAIR facility at GSI \cite{FAIR}.

\section{\label{sec5}Conclusions and outlook}

In this work the conventional linear Walecka model of relativistic 
hadrodynamics was extended by introducing non-linear derivative operators 
in the interaction Lagrangian. A generalization of standard field theoretical 
technics for Lagrangians depending on all higher order derivatives of a 
field was necessary in deriving field equations and applying the 
Noether theorem in a consistent way. Starting from rather complicated field 
equations it was possible to reduce them to compact and simple expressions 
by resuming all higher order derivatives.

The model was applied in nuclear matter using a single parameter of natural 
hadronic mass scale. Astonishing were the non-linear density effects on the 
vector field without the introduction of any additional selfinteraction terms 
in the original Lagrangian of the linear Walecka model. These effects lead 
to a softening of the equation of state for nuclear and pure neutron matter at 
high densities. It was possible to describe quantitatively well the empirically 
known saturation properties. The results were also comparable to predictions 
of microscopic DB calculations over a wide density range. 
The approach lead furthermore to a momentum dependence of the selfenergies. 
Most noticeable was the correct reproduction of the energy dependence of the 
Schr\"{o}dinger equivalent optical potential by utilizing the same parameter. 
The microscopic origin or the underlying mechanism of the desired results 
is not relevant at the present level of study. 

In principle, one could go beyond the present study and apply the model 
to more complex nuclear systems, e.g., finite nuclei. In this context 
the present model can be further extended by introducing different cut-off 
parameters or different non-linear operator functions than the 
exponential form used in this work. Another physical object of interest would 
be heavy ion collisions at high energies with highest baryon 
compression, as they are planned at the new FAIR facility at GSI, in 
which the density and, in particular, the energy dependence of the nuclear 
mean-field play an important role. The application of the present model to 
heavy-ion collisions in the spirit of a covariant transport theory based 
on the present Lagrangian would be a great challenge for the future in 
studying hadronic matter under extreme conditions with the ultimative goal 
of exploring the equation of state at supra-normal densities.

{\it Acknowledgments:}
This work is supported by BMBF. 


\begin{appendix}
{\bf Appendix}
\vspace{-0.7cm}
\section{\label{app1}
Generalized Euler-Lagrange equations and Noether theorem}

The procedure to derive the generalized Euler-Lagrange equations 
follows the canonical formalism and starts from the variation of 
the action. Standard steps like the integration by parts lead to 
the following expression:

\begin{eqnarray}
\frac{\partial{\cal L}}{\partial\phi}
-
 \partial_{\alpha}\frac{\partial{\cal L}}{\partial(\partial_{\alpha}\phi)}
+
 \partial_{\alpha}\partial_{\beta}\frac{\partial{\cal L}}{\partial(\partial_{\alpha}\partial_{\beta}\phi)}
&& + \cdots  +  \label{Psi-2}\\
&& (-)^{n}\partial_{\alpha_{1}}\partial_{\alpha_{2}}\cdots\partial_{\alpha_{n}}
\frac{\partial{\cal L}}
{\partial(\partial^{\alpha_{1}}\partial^{\alpha_{2}}\cdots\partial^{\alpha_{n}})}=
0 \nonumber
\;. 
\end{eqnarray}

The derivation of the conserved energy-momentum tensor and current is more 
involved. 
The conserved energy-momentum tensor and current are obtained from 
a variation of the Lagrangian with respect to infinitesimal changes 
of the field $\phi$ and its argument. The total variation of the field reads
\begin{equation}
\delta_{T}\phi(x) :=  \phi^{\prime}(x^{\prime})-\phi(x) 
=
\left[ 
	\phi^{\prime}(x^{\prime}) - \phi(x^{\prime})
\right] 
+
\left[ 
	\phi(x^{\prime}) - \phi(x)
\right] 
\label{var1}
\end{equation}
which in first order reduces to the usual form
\begin{equation}
\delta_{T}\phi(x) = \delta\phi(x) + \partial_{\alpha}\phi(x)\delta x^{\alpha}
\label{var2}
\end{equation}
with $\delta\phi=\phi^{\prime}-\phi$ and $\delta x^{\mu}$ being the small variation of 
the field at fixed argument and the infinitesimal change of it, respectively. 

The variation of the Lagrangian density functional thus reads
\begin{eqnarray}
& & 
{\cal L}[\phi^{\prime}(x^{\prime}),
\partial_{\alpha}\phi^{\prime}(x^{\prime}), 
\partial_{\alpha}\partial_{\beta}\phi^{\prime}(x^{\prime}),\cdots,]
-
{\cal L}[\phi(x^{\prime}),
\partial_{\alpha}\phi(x^{\prime}), 
\partial_{\alpha}\partial_{\beta}\phi(x^{\prime}),\cdots,]
\nonumber\\
& & +
{\cal L}[\phi(x^{\prime}),
\partial_{\alpha}\phi(x^{\prime}), 
\partial_{\alpha}\partial_{\beta}\phi(x^{\prime}),\cdots,]
-
{\cal L}[\phi(x),
\partial_{\alpha}\phi(x), 
\partial_{\alpha}\partial_{\beta}\phi(x),\cdots,] = 0
\;. \label{var3} \nonumber \\
\end{eqnarray}
The first line in Eq. (\ref{var3}) is just the variation of ${\cal L}$ with 
respect to the fields $\phi(x),\partial_{\alpha}\phi,\cdots$ at fixed argument, 
$\delta{\cal L}$, whereas the second line gives the variation of the 
Lagrangian with respect to the argument which can be evaluated up to first 
order. In total we obtain:
\begin{equation}
\delta{\cal L} + \partial_{\alpha}{\cal L}\delta x^{\alpha} = 0
\label{var4} 
\end{equation}
with 
\begin{equation}
\delta{\cal L} = 
 \frac{\partial{\cal L}}{\partial\phi} \delta\phi
+
 \frac{\partial{\cal L}}{\partial(\partial_{\alpha}\phi)} 
 \delta(\partial_{\alpha}\phi)
+
 \frac{\partial{\cal L}}{\partial(\partial_{\alpha}\partial_{\beta}\phi)} 
 \delta (\partial_{\alpha}\partial_{\beta}\phi)
+
\cdots
\;. \label{var5} 
\end{equation}

Inserting in the first term in Eq. (\ref{var5}) the Euler-Lagrange 
equations (\ref{Euler}), we obtain after the application of the 
product rule following expression:
\begin{eqnarray}
& & 
\delta{\cal L}  = 
\partial_{\mu}\left\{
  \left[
    \frac{\partial{\cal L}}{\partial(\partial_{\mu}\phi)}
  - \partial_{\beta}
    \frac{\partial{\cal L}}{\partial(\partial_{\mu}\partial_{\beta}\phi)}
  + \partial_{\beta}\partial_{\gamma}
    \frac{\partial{\cal L}}{\partial(\partial_{\mu}\partial_{\beta}\partial_{\gamma}\phi)}
  \mp \cdots
  \right]\delta\phi
\right .
\nonumber\\
& & 
 \left . 
\hspace*{1.2cm} + \left[
    \frac{\partial{\cal L}}{\partial(\partial_{\mu}\partial_{\nu}\phi)}
  - \partial_{\gamma}
    \frac{\partial{\cal L}}{\partial(\partial_{\mu}\partial_{\beta}\partial_{\nu}\phi)}
  \pm \cdots
  \right]\partial_{\nu}\delta\phi
\right .
\nonumber\\
& & 
 \left . 
\hspace*{1.2cm} + \left[
    \frac{\partial{\cal L}}{\partial(\partial_{\mu}\partial_{\nu}\partial_{\xi}\phi)}
  - \partial_{\gamma}
    \frac{\partial{\cal L}}{\partial(\partial_{\mu}\partial_{\gamma}\partial_{\nu}\partial_{\xi}\phi)}
  \pm \cdots
  \right]\partial_{\nu}\partial_{\xi}\delta\phi 
\right .
\nonumber\\
& & 
 \left . 
\hspace*{1.2cm} + \cdots
\right \}
\;. \label{delta_L}
\end{eqnarray}

As next step we insert the expression \ref{delta_L} into (\ref{var4}) obtaining: 
\begin{equation}
\partial_{\alpha}
\left( 
  J^{\alpha} - T^{\alpha\nu}\delta x_{\nu} 
\right) = 0
\label{varf} 
\end{equation}
which is the desired result, see also (\ref{Noether-a}).

\section{\label{app2}Field equations in the NLD model}
In this appendix the most important steps leading to the field equations 
for the spinor $\Psi$ and the meson fields $\sigma$ and $\omega$ are given. 
The starting basis for the derivation of a Dirac equation from the 
Lagrangian density (\ref{NDC}) are the generalized Euler-Lagrange equations 
(\ref{Euler}). We give the terms up to second order explicitly. 

In order to show the disappearance of the rearrangement terms in 
nuclear matter, we choose in the derivation of the Dirac equation for the auxiliary 
unit vector $v^{\mu}$ the usual nucleon current, see Eq. (\ref{unitvector}), 
in which rearrangement terms only in the vector selfenergy will appear. 
For the derivation of the Proca 
equation ($\omega$ meson) $v^{\mu}=\omega^{\mu}/\sqrt{\omega_{\alpha}\omega^{\alpha}}$ 
will be chosen, in order to show that the dissapereance of the rearrangement terms in 
nuclear matter is of general nature, i.e., independent of the particular choice of the 
auxiliary vector $v^{\mu}$.

\subsection{\label{app2a}Dirac equation}
For the derivation of the Dirac equation we start from the generalized 
Euler-Lagrange equations (\ref{Psi-2}). We derive explicitely the first three 
orders in the derivative expansion:
\vspace{-0.5cm}
\begin{eqnarray}
& & 
\frac{\partial {\cal L}}{\partial\overline{\Psi}} +
\frac{\partial {\cal L}}{\partial v_{\mu}}\frac{\partial v^{\mu}}{\partial\overline{\Psi}} = 
\frac{1}{2}\gamma_{\mu}i\partial^{\mu}\psi - m\Psi 
\nonumber\\
& & - \frac{1}{2}g_{\omega}
\left[
	\gamma_{\mu}\Psi\omega^{\mu}+\gamma_{\mu}\evr\Psi\omega^{\mu}
\right]
+  \frac{1}{2}g_{\sigma}
\left[
	\Psi\sigma+\evr\Psi\sigma
\right]
\nonumber\\
& - & 
\left[
\frac{1}{2}\Gv 
\left(
	\overline{\Psi}
	(i\stackrel{\leftarrow}{\partial_{\mu}})\evl\gamma^{\alpha}\Psi\omega_{\alpha}
	+
	\overline{\Psi}
	\gamma^{\alpha}\evr (-i\stackrel{\rightarrow}{\partial_{\mu}})
	\Psi\omega_{\alpha}
\right)
\right.
\nonumber\\
& & 
\left.
- \frac{1}{2}\Gs
\left(
	\overline{\Psi}
	(i\stackrel{\leftarrow}{\partial_{\mu}})\evl\Psi\sigma
	+
	\overline{\Psi}
	\evr (-i\stackrel{\rightarrow}{\partial_{\mu}})
	\Psi\sigma
\right)
\right]
\nonumber\\
& & \times
\left[
\frac{\gamma^{\mu}\Psi}{\sqrt{j_{\alpha}j^{\alpha}}} - 
\frac{j^{\mu}j_{\nu}\gamma^{\nu}\Psi}{\sqrt{j_{\alpha}j^{\alpha}}^{3}}
\right]
\nonumber\\
& & 
\frac{\partial {\cal L}}{\partial(\partial_{\mu}\overline{\Psi})} = 
- \frac{1}{2}i\gamma^{\mu}\psi
- \frac{1}{2}\Gv v^{\mu} i \gamma_{\alpha}\Psi\omega^{\alpha}
+ \frac{1}{2}\Gs v^{\mu} i \Psi\sigma
\nonumber\\
& & 
\frac{\partial {\cal L}}{\partial(\partial_{\mu}\partial_{\nu}\overline{\Psi})} = 
- \frac{1}{2}\Gvb \frac{1}{2!} v^{\mu}v^{\nu} ii \gamma_{\alpha}\Psi\omega^{\alpha}
+ \frac{1}{2}\Gsb \frac{1}{2!} v^{\mu}v^{\nu} ii \Psi\sigma
\;. \label{Psi-1}
\end{eqnarray}

Inserting (\ref{Psi-1}) into (\ref{Psi-2}) we obtain after some trivial algebra 
(up to derivatives of the meson fields which in any case vanish in nuclear matter):
\begin{eqnarray}
& & 
\left( 
	\gamma_{\mu}i\partial^{\mu} - m
\right)\Psi 
- \frac{1}{2}g_{\omega}\gamma_{\mu}\evr \Psi \omega^{\mu}
+ \frac{1}{2}g_{\sigma}\evr \Psi \sigma
\nonumber\\
& & 
- \frac{1}{2}g_{\omega}\gamma_{\mu}
\left( 
	1 - \frac{1}{\Lambda}v_{\alpha}i\partial^{\alpha}
	+ \frac{1}{2!}\frac{1}{\Lambda^{2}}v_{\alpha}v_{\beta}
		i\partial^{\alpha}i\partial^{\beta}
	\, \mp \, \cdots
\right)\Psi\omega^{\mu}
\nonumber\\
& & 
+ \frac{1}{2}g_{\sigma}
\left( 
	1 - \frac{1}{\Lambda}v_{\alpha}i\partial^{\alpha}
	+ \frac{1}{2!}\frac{1}{\Lambda^{2}}v_{\alpha}v_{\beta}
		i\partial^{\alpha}i\partial^{\beta}
	\, \mp\, \cdots
\right)\Psi\sigma + \Sigma_{r}\Psi
= 0
\;. \label{Psi-3}
\end{eqnarray}

with the rearrangement contribution: 
\begin{eqnarray}
& & 
\Sigma_{r} = 
\left[
- \frac{1}{2}\Gv 
\left(
	\overline{\Psi}
	(i\stackrel{\leftarrow}{\partial_{\mu}})\evl\gamma^{\delta}\Psi\omega_{\delta}
	+
	\overline{\Psi}
	\gamma^{\delta}\evr (-i\stackrel{\rightarrow}{\partial_{\mu}})
	\Psi\omega_{\delta}
\right)
\right.
\nonumber\\
& & 
\left.
+ \frac{1}{2}\Gs
\left(
	\overline{\Psi}
	(i\stackrel{\leftarrow}{\partial_{\mu}})\evl\Psi\sigma
	+
	\overline{\Psi}
	\evr (-i\stackrel{\rightarrow}{\partial_{\mu}})
	\Psi\sigma
\right)
\right]
\nonumber\\
& & \times
\left[
\frac{\gamma^{\mu}}{\sqrt{j_{\alpha}j^{\alpha}}} - 
\frac{j^{\mu}j_{\nu}\gamma^{\nu}}{\sqrt{j_{\alpha}j^{\alpha}}^{3}}
\right]
\;. \label{rearr_bar}
\end{eqnarray}

Note that, the derivatives inside the brackets act to the spinors on the right. 
The terms inside the brackets represent Taylor expansions which can be resumed 
to the exponential function $\evr$. This leads us to the following Dirac equation
\begin{equation}
\left[
	\gamma_{\mu}(i\partial^{\mu}-\Sigma^{\mu}) - 
	(m-\Sigma_{s})
\right]\Psi = 0
\;, 
\label{Psi-4}
\end{equation}
as it was written in Eq. (\ref{Dirac}).

In infinite nuclear matter only the $0$-component of the $\omega$ meson field 
survives, and in the mean-field approximation all density operators are replaced 
by expectation values. Furthermore only the $0$-component of the usual nucleon 
current $j^{\mu}=\langle \overline{\Psi}\gamma^{\mu}\Psi\rangle$ remains. 
Thus, the rearrangement contribution reads:
\begin{equation}
\Sigma_{r} =  
\left[
\Gv \langle \overline{\Psi}p_{\mu}\evnm \gamma^{0}\Psi\rangle \omega_{0}
- \Gs \langle \overline{\Psi}p_{\mu}\evnm \Psi\rangle \sigma
\right]
\left[
\frac{\gamma^{\mu}}{\sqrt{j_{\alpha}j^{\alpha}}} - 
\frac{j^{\mu}j_{\nu}\gamma^{\nu}}{\sqrt{j_{\alpha}j^{\alpha}}^{3}}
\right]
\;, \nonumber
\end{equation}
and by evaluating the contraction between the expressions inside the 
two brackets we obtain:
\begin{eqnarray}
\Sigma_{r} & = & 
\left[
\Gv \langle \overline{\Psi}p_{0}\evnm \gamma^{0}\Psi\rangle \omega_{0}
- \Gs \langle \overline{\Psi}p_{0}\evnm \Psi\rangle \sigma
\right]
\left[
\frac{\gamma^{0}}{j^{0}} - \frac{\gamma^{0}}{j^{0}}
\right]
\nonumber\\
& - & \left[
\Gv \langle \overline{\Psi}\vec{p}\evnm \gamma^{0}\Psi\rangle \omega_{0}
- \Gs \langle \overline{\Psi}\vec{p}\evnm \Psi\rangle \sigma
\right]
\frac{\vec{\gamma}}{j^{0}} = 0
\;. \label{rearr_bar_NM}
\end{eqnarray}
The first term in Eq. (\ref{rearr_bar_NM}) is trivially zero. In 
the second line of Eq. (\ref{rearr_bar_NM}) both expectation values 
are odd functions of the vector $\vec{p}$ and therefore they vanish.

\subsection{\label{app2b}Meson field equations of motion}
For the meson fields $\sigma$ and $\omega$ the derivations are straightforward, 
since here one has to use the standard Euler-Lagrange equations. The following 
Proca equation which includes the rearrangement and the source terms is 
readily derived:
\begin{equation}
\partial_{\mu}F^{\mu\nu} + m_{\omega}^{2}\omega^{\nu} = 
\frac{1}{2}g_{\omega}
\left[
	\overline{\Psi}\evl \gamma^{\nu}\Psi + \overline{\Psi}\gamma^{\nu}\evr \Psi
\right] + \Omega^{\nu}
\end{equation}
The rearrangement term in the Proca equation appears if for the auxiarily vector 
$v^{\mu}=\omega^{\mu}/\sqrt{\omega_{\alpha}\omega^{\alpha}}$ is chosen. It takes 
the following form:
\begin{eqnarray}
& & 
\Omega^{\nu} := 
- \, \frac{1}{2}\frac{g_{\omega}}{\Lambda}
\left[
	\overline{\Psi}
	\evl
	\left( 
		  \frac{\stackrel{\leftarrow}{i\partial}^{\nu}}{\sqrt{\omega_{\alpha}\omega^{\alpha}}}
		- \frac{\stackrel{\leftarrow}{i\partial}^{\alpha}\omega_{\alpha}\omega^{\nu}}
			{\sqrt{\omega_{\alpha}\omega^{\alpha}}^{3}}
	\right)\gamma^{\delta}\Psi\omega_{\delta}
\right.
\nonumber\\
& &
\left. 
\hspace{2.2cm}	
+ \, 
	\overline{\Psi}
	\left( 
		-  \frac{\stackrel{\rightarrow}{i\partial}^{\nu}}{\sqrt{\omega_{\alpha}\omega^{\alpha}}}
		+ \frac{\omega_{\alpha}\omega^{\nu}\stackrel{\rightarrow}{i\partial}^{\alpha}}
			{\sqrt{\omega_{\alpha}\omega^{\alpha}}^{3}}
	\right)\evr\gamma^{\delta}\Psi\omega_{\delta}
\right]
\nonumber\\
& & 
\hspace{1.2cm} 
+ \, \frac{1}{2}\frac{g_{\sigma}}{\Lambda}
\left[
	\overline{\Psi}
	\evl
	\left( 
		  \frac{\stackrel{\leftarrow}{i\partial}^{\nu}}{\sqrt{\omega_{\alpha}\omega^{\alpha}}}
		- \frac{\stackrel{\leftarrow}{i\partial}^{\alpha}\omega_{\alpha}\omega^{\nu}}
			{\sqrt{\omega_{\alpha}\omega^{\alpha}}^{3}}
	\right)\Psi\sigma
\right.
\nonumber\\
& & 
\left. 
\hspace{2.2cm}
	+ \, \overline{\Psi}
	\left( 
		-  \frac{\stackrel{\rightarrow}{i\partial}^{\nu}}{\sqrt{\omega_{\alpha}\omega^{\alpha}}}
		+ \frac{\omega_{\alpha}\omega^{\nu}\stackrel{\rightarrow}{i\partial}^{\alpha}}
			{\sqrt{\omega_{\alpha}\omega^{\alpha}}^{3}}
	\right)\evr\Psi\sigma
\right]
\label{rearr_o}
\end{eqnarray}
In nuclear matter and in the mean-field approximation the rearrangement contribution 
$\Omega^{\mu}$ to the vector meson vanishes for the same reasons as for that rearrangement 
term for the nucleon field (see previous subsection). 

\section{\label{app3}Derivation of the Noether current}
Here the most important steps leading to the Noether current 
(\ref{rhoBar}) starting from the general expression (\ref{Noether-Current}) are explicitly 
shown. The derivation of the energy-momentum tensor (\ref{Noether}) is 
then straightforward. We show that all higher order derivative terms can be 
resumed to simple exponential expressions for nuclear matter at rest in the 
mean-field approximation.

The derivative of the Lagrangian density (\ref{NDC}) with respect to 
higher order partial derivatives has been demonstrated in 
the appendix \ref{app2a}, see in particular Eq. (\ref{Psi-1}), thus we don't repeat 
this intermediate step.

We consider the Noether current in the NLD model starting from the general 
expression (\ref{Noether-Current}) and taking only terms up to third order. With 
$\delta\bar{\Psi}=-i\epsilon\bar{\Psi}$ and 
$\delta \Psi=i\epsilon\Psi$ we obtain:
\begin{eqnarray}
\frac{J^{\mu}}{i\epsilon} & = &
  \left[
    \frac{\partial{\cal L}}{\partial(\partial_{\mu}\Psi)}
  - \partial_{\beta}
    \frac{\partial{\cal L}}{\partial(\partial_{\mu}\partial_{\beta}\Psi)}
  + \partial_{\beta}\partial_{\gamma}
    \frac{\partial{\cal L}}{\partial(\partial_{\mu}\partial_{\beta}\partial_{\gamma}\Psi)}
  \right]\Psi
\nonumber\\
& + & \left[
    \frac{\partial{\cal L}}{\partial(\partial_{\mu}\partial_{\beta}\Psi)}
  - \partial_{\gamma}
    \frac{\partial{\cal L}}{\partial(\partial_{\mu}\partial_{\beta}\partial_{\gamma}\Psi)}
  \right]\partial_{\beta}\Psi
\nonumber\\
& + & 
\left[
    \frac{\partial{\cal L}}{\partial(\partial_{\mu}\partial_{\nu}\partial_{\xi}\Psi)}
  - \partial_{\gamma}
\frac{\partial{\cal L}}{\partial(\partial_{\mu}\partial_{\gamma}\partial_{\nu}\partial_{\xi}\Psi)}
  \right]\partial_{\nu}\partial_{\xi}\Psi
\nonumber\\
& - & 
	\bar{\Psi}
  \left[
    \frac{\partial{\cal L}}{\partial(\partial_{\mu}\bar{\Psi})}
  - \partial_{\beta}
    \frac{\partial{\cal L}}{\partial(\partial_{\mu}\partial_{\beta}\bar{\Psi})}
  + \partial_{\beta}\partial_{\gamma}
    \frac{\partial{\cal L}}{\partial(\partial_{\mu}\partial_{\beta}\partial_{\gamma}\bar{\Psi})}
  \right]
\nonumber\\
& + &
		\partial_{\beta}\bar{\Psi} 
		\left[
    \frac{\partial{\cal L}}{\partial(\partial_{\mu}\partial_{\beta}\bar{\Psi})}
  - \partial_{\gamma}
    \frac{\partial{\cal L}}{\partial(\partial_{\mu}\partial_{\beta}\partial_{\gamma}\bar{\Psi})}
  \right]
\nonumber\\
& + & \partial_{\nu}\partial_{\xi}\bar{\Psi}\left[
    \frac{\partial{\cal L}}{\partial(\partial_{\mu}\partial_{\nu}\partial_{\xi}\bar{\Psi})}
  - \partial_{\gamma}
    \frac{\partial{\cal L}}{\partial(\partial_{\mu}\partial_{\gamma}\partial_{\nu}\partial_{\xi}\bar{\Psi})}
  \right]
\;. \label{strom-1}
\end{eqnarray}
We rewrite now Eq. (\ref{strom-1}) by separating the terms between the different orders 
in the partial derivatives:
\begin{equation}
\frac{J^{\mu}}{i\epsilon} = {\cal O}^{(0)} + {\cal O}^{(1)} + {\cal O}^{(2)} 
+ \cdots + {\cal O}^{(n)}
\end{equation}
with the different contributions given by:
\begin{eqnarray}
{\cal O}^{(0)} =
\left( 
	\frac{\partial{\cal L}}{\partial(\partial_{\mu}\Psi)}\Psi
	-
	\bar{\Psi}\frac{\partial{\cal L}}{\partial(\partial_{\mu}\bar{\Psi})}
\right)
\;, \label{0-order}
\end{eqnarray}

\begin{eqnarray}
{\cal O}^{(1)} &=& - 
\left( 
	\partial_{\beta}
	\frac{\partial{\cal L}}{\partial(\partial_{\mu}\partial_{\beta}\Psi)}
	\Psi
	-
	\bar{\Psi}
	\partial_{\beta}
	\frac{\partial{\cal L}}{\partial(\partial_{\mu}\partial_{\beta}\bar{\Psi})}
\right) \nonumber \\
&& +
\left( 
	\frac{\partial{\cal L}}{\partial(\partial_{\mu}\partial_{\beta}\Psi)}
	\partial_{\beta}\Psi
	-
	\partial_{\beta}\bar{\Psi}
	\frac{\partial{\cal L}}{\partial(\partial_{\mu}\partial_{\beta}\bar{\Psi})}
\right)
\;, \label{1-order}
\end{eqnarray}

\begin{eqnarray}
{\cal O}^{(2)} & = & + 
\left( 
	\partial_{\beta}\partial_{\gamma}
	\frac{\partial{\cal L}}{\partial(\partial_{\mu}\partial_{\beta}\partial_{\gamma}\Psi)}
	\Psi
	-
	\bar{\Psi}
	\partial_{\beta}\partial_{\gamma}
	\frac{\partial{\cal L}}{\partial(\partial_{\mu}\partial_{\beta}\partial_{\gamma}\bar{\Psi})}
\right)
\nonumber\\
&& - 
\left( 
	\partial_{\gamma}
	\frac{\partial{\cal L}}{\partial(\partial_{\mu}\partial_{\beta}\partial_{\gamma}\Psi)}
	\partial_{\beta}\Psi
	-
	\partial_{\beta}\bar{\Psi}
	\partial_{\gamma}
	\frac{\partial{\cal L}}{\partial(\partial_{\mu}\partial_{\beta}\partial_{\gamma}\bar{\Psi})}
\right)
\nonumber\\
&& + 
\left( 
	\frac{\partial{\cal L}}{\partial(\partial_{\mu}\partial_{\beta}\partial_{\gamma}\Psi)}
	\partial_{\beta}\partial_{\gamma}\Psi
	-
	\partial_{\beta}\partial_{\gamma}\bar{\Psi}
	\frac{\partial{\cal L}}{\partial(\partial_{\mu}\partial_{\beta}\partial_{\gamma}\bar{\Psi})}
\right)
\;. \label{2-order}
\end{eqnarray}

In mean-field approximation it turns out that first order terms (in energy) 
appear twice, second order terms appear three times, and so forth:
\begin{equation}
{\cal O}^{(0)} =  
\langle \bar{\Psi}\gamma^{0}\Psi \rangle
+ \Gv \langle\bar{\Psi}\gamma^{0}\Psi\rangle \omega_{0} 
- \Gs \langle\bar{\Psi}\Psi\rangle \sigma,
\end{equation}
\begin{equation}
{\cal O}^{(1)} =  
-\Gv \frac{2}{2!} \langle\bar{\Psi}\gamma^{0}\frac{E}{\Lambda}\Psi\rangle \omega_{0}
+
\Gs \frac{2}{2!} \langle\bar{\Psi}\frac{E}{\Lambda}\Psi\rangle \sigma,
\end{equation}
\begin{equation}
{\cal O}^{(2)} = 
\Gv \frac{3}{3!} \langle\bar{\Psi}\gamma^{0}\frac{E^{2}}{\Lambda^{2}}\Psi\rangle \omega_{0}
-
\Gs \frac{3}{3!} \langle\bar{\Psi}\frac{E^{2}}{\Lambda^{2}}\Psi\rangle \sigma
\;, \label{step1}
\end{equation}
and for the $n-th$-order one obtains
\begin{equation}
{\cal O}^{(n-1)} = 
(-)^{n}\Gv \frac{n}{n!} \langle\bar{\Psi}\gamma^{0}\frac{E^{n-1}}{\Lambda^{n-1}}\Psi\rangle \omega_{0}
-
(-)^{n}\Gs \frac{n}{n!} \langle\bar{\Psi}\frac{E^{n-1}}{\Lambda^{n-1}}\Psi\rangle \sigma
\;. \label{stepn}
\end{equation}
Therefore all the infinite series of terms can be resumed into simple expressions. 
Adding all these terms and considering the fact that $\frac{n}{n!}=\frac{1}{(n-1)!}$ 
we obtain the desired result, see Eq.~(\ref{rhoBar}).

\end{appendix}

\newpage



\begin{thebibliography}{100}

\bibitem{QHD97}
B.~Serot and J.D.~Walecka, 
Int.\ J.\ Mod.\ Phys.\ {\bf E6} (1997) 631.

\bibitem{hama}
E.D.~Cooper, et al., 
Phys.\ Rev.\ {\bf C47} (1993) 297.

\bibitem{duerr}
H.P.~Duerr, 
Phys.\ Rev.\ {\bf 103} (1956) 469.

\bibitem{wal74}
J.D.~Walecka, 
Ann.\ Phys.\ (NY) {\bf 83} (1974) 491.

\bibitem{DB}
R.~Brockmann and R.~Machleidt, 
Phys.\ Rev.\ {\bf C42} (1990) 1965.

\bibitem{boguta}
J.~Boguta, A.R.~Bodmer, Nucl.\ Phys.\ {\bf A292} (1977) 413.

\bibitem{sugahara}
Y.~Sugahara and H.~Toki, Nucl.\ Phys.\ {\bf A579} (1994) 557.

\bibitem{Toki_DDH}
R.~Brockmann and H.~Toki, Phys.\ Rev.\ Lett.\ {\bf 68} (1992) 3408.

\bibitem{DDH}
C.~Fuchs, H.~Lenske and H.H.~Wolter, 
Phys.\ Rev.\ {\bf C52} (1995) 3043.

\bibitem{cass}
B.~Bl\"{a}ttel, V.~Koch and U.~Mosel, 
Rep.\ Prog.\ Phys.\ {\bf 56} (1993) 1.

\bibitem{ZM}
J.~Zimanyi and S.A.~Moszkowski, 
Phys.\ Rev.\ {\bf C42} (1990) 1416.

\bibitem{DC}
S.~Typel, Phys.\ Rev.\ {\bf C71} (2005) 064301.

\bibitem{DB-e}
B.~ter~Haar and R.~Malfliet, 
Phys.\ Rev.\ {\bf C36} (1987) 1611;\\
B.~ter~Haar and R.~Malfliet, Phys.\ Rep.\ {\bf 149} (1987) 207.



\bibitem{FAIR}
P.~Senger, Phys.\ Part.\ Nucl.\ {\bf 39} (2008) 1055.

\end{thebibliography}
\end{document}